\documentclass[final,3p,times,twocolumn]{elsarticle}

\usepackage{graphicx}
\usepackage{amsmath,bm}
\usepackage{amssymb}
\usepackage{color,soul}
\usepackage{xcolor}

\journal{Chaos, Solitons \& Fractals}

\begin{document}

\begin{frontmatter}

\title{Probabilistic Memristive Networks: Application of a Master Equation to Networks of Binary ReRAM cells}

\author[AD1]{Vincent~J.~Dowling}

\author[AD2]{Valeriy~A.~Slipko}

\author[AD1]{Yuriy~V.~Pershin\corref{cor1}}
\ead{pershin@physics.sc.edu}

\cortext[cor1]{Corresponding author}

\address[AD1]{Department of Physics and Astronomy, University of South Carolina, Columbia, SC 29208 USA}
\address[AD2]{Institute of Physics, Opole University, Opole 45-052, Poland}

\begin{abstract}
The possibility of using non-deterministic circuit components has been gaining significant attention in recent years. The modeling and simulation of their circuits require novel approaches, as now the state of a circuit at an arbitrary moment in time cannot be precisely predicted. Generally, these circuits should be described in terms of probabilities, the circuit variables should be calculated on average, and correlation functions should be used to explore interrelations among the variables. In this paper, we use, for the first time, a master equation to analyze the networks composed of probabilistic binary memristors. Analytical solutions of the master equation for the case of identical memristors connected in-series and in-parallel are found. Our analytical results are supplemented by results of numerical simulations that extend our findings beyond the case of identical memristors. The approach proposed in this paper facilitates the development of probabilistic/stochastic electronic circuits and advance their real-world applications.
\end{abstract}


\end{frontmatter}


\section{Introduction}
Resistance switching memories  are a very promising class of memory devices that have been intensively studied in the past few decades. The simple device structure, scalability, fast speed, and compatibility with current silicon technology make them ideal candidates for the next generation of storage-class memory~\cite{Burr08a}. However, significant temporal (cycle to cycle) and spacial (device to device) parameter fluctuations observed in all reported ReRAM cells~\cite{Yu12a} present a major obstacle for their wide-scale commercialization. As it is obvious that the stochasticity is an inherent feature of the resistance switching memories, the accurate and predictable modeling of single ReRAM devices and circuits thereof require approaches beyond the deterministic models (such as in Refs.~\cite{Strachan13a,Kvatinsky15a,Panda2018,la2019compact,lee2020quantitative}).

The method of stochastic differential equations~\cite{oksendal2013stochastic} is the standard way to take account for
fluctuations in otherwise deterministic models. Some applications of this method to the problem of stochasticity in ReRAM cells  have been reported~\cite{Stotland12a,Naous16a,Patterson_2016,Gough17a} including the postulation of stochastic memory elements by YVP and Di Ventra in 2011~\cite{pershin11a}. However, the method of stochastic differential equations has yet to be adopted widely in the ReRAM community, possibly because of its relative complexity. Menzel et al.~\cite{menzel2014statistical} developed a kinetic Monte Carlo model for the resistive switching in ECM cells describing the filament growth on a single-atom level.

The randomness in the ReRAM switching can be described in terms of probabilities ignoring the details of microscopic dynamics. In particular, it was shown experimentally that the off-to-on transition in electrochemical metallization (ECM) cells occurs according to the Poisson distribution~\cite{jo2009programmable,gaba2013stochastic,gaba2014memristive}. Moreover,
Medeiros-Ribeiro et al.~\cite{Medeiros_Ribeiro_2011} investigated the distribution of switching times in TiO$_2$ valence change memories, which are another type of ReRAM cells. They found that both off-to-on and on-to-off transitions are described by a log-normal distribution. The Poisson distribution observed in ECM cells~\cite{jo2009programmable,gaba2013stochastic,gaba2014memristive} indicates a Markovian dynamics that can be conveniently described in terms of a master equation.

In this paper we consider networks composed of $N$ binary ReRAM cells, or simply memristors\footnote{The claim~\cite{chua2011a} that ReRAM cells are memristors~\cite{chua71a} is debatable~\cite{Kim20a}.}, governed by Poisson switching statistics. A master equation is introduced to describe
the network dynamics on average that, in a particular realization, consists in consecutive jumps over some of $2^N$ states.
Generally, the master equation can be used to describe networks with non-identical devices. However, the problem complexity is significantly reduced in the case of identical devices.
In this paper, the master equation is solved analytically for the cases of identical memristors connected in-series and in-parallel. The derivations made in this work assume an abstract two-state model of ReRAM cells supported by experiments~\cite{jo2009programmable,gaba2013stochastic,gaba2014memristive} and could be verified with those devices that behave according to such a model.

This paper is organized as follows. In Sec.~\ref{sec:2} preliminaries are presented that include
the probabilistic model summary, and numerical simulation details. Sec.~\ref{sec:3} presents the master equation, and its solutions for the cases of in-series and in-parallel
connected memristors. Correlation functions are introduced and derived in Sec.~\ref{sec:4}. We conclude in Sec.~\ref{sec:5}.
The appendix contains a concise mathematical treatment of the dynamics of the off-to-on transition in the circuit of $N$ in-parallel and in-series
connected memristors, and some other supplementary results.

\section{Preliminaries} \label{sec:2}

\subsection{ReRAM cell switching model}

In this paper we consider the networks of probabilistic binary resistance switching memories. By binary~\cite{suri2013bio,Truong2014} we mean that our
devices can be found in one of two well-defined resistance states, $R_{on}$ and $R_{off}$. By probabilistic
we mean that randomness plays a role in the process of switching.
It is assumed that the switching events are instantaneous, and their probability is a well-known function of the applied voltage or current. For compactness, we use the terms ``memristors''
and ``probabilistic binary resistance switching memories'' interchangeably. However, it should be emphasized that
the devices considered here are not described by the memristor equations~\cite{chua71a,chua76a}.

\begin{figure}[tb]
  \centering
     \includegraphics[width=0.98\columnwidth]{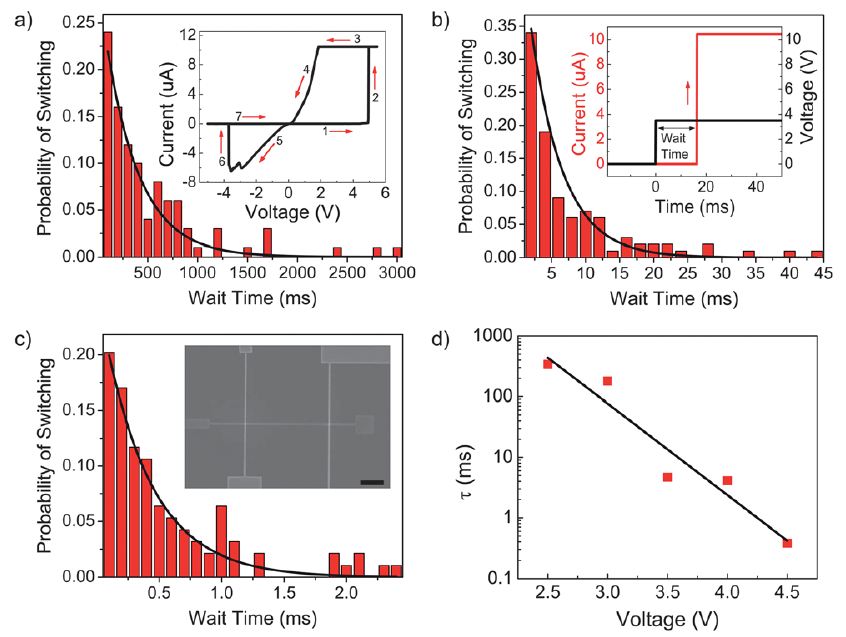}
  \caption{ (a-c) Experimentally measured distributions of switching (wait) times in Ag-SiO$_2$ cells~\cite{gaba2013stochastic}. (d) Voltage-dependence of the characteristic switching time $\tau(V)$ (see Eq.~(\ref{eq:tau})). The solid lines in (a-c) are fitting to the Poisson distribution (Eq.~(\ref{eq:Poisson})).
  An example of $I-V$ curve, switching time measurement, and sample micrograph are presented in the insets in (a)-(c), respectively. Reprinted with permission from~\cite{gaba2013stochastic}.}\label{fig:0}
\end{figure}

The studies~\cite{jo2009programmable,gaba2013stochastic,gaba2014memristive} of the off-to-on transition in electrochemical metallization cells have shown that the probability of switching within a small time interval $\Delta t \ll \tau(V)$ follows the Poisson distribution
\begin{equation}\label{eq:Poisson}
  P(t)=\frac{\Delta t}{\tau(V)} e^{-t/\tau(V)},
\end{equation}
where $\tau(V)$ is the voltage-dependent characteristic switching time, and $V$ is the voltage across the cell.
Fig.~\ref{fig:0} presents results of experimental measurements reprinted from Ref.~\cite{gaba2013stochastic}. In these experiments, a single memristor initialized into the off-state  is subjected to a constant voltage starting at $t=0$. The time of the transition from the off- into the on-state is traced by a step in the current (see the inset in Fig.~(\ref{fig:0})). Eq.~(\ref{eq:Poisson}) was used to fit the distributions of the switching times. Technically, it describes the probability of switching within the time interval from $t$ to $t+\Delta t$ for a memristor in the off-state at $t=0$.

Moreover, a very good agreement with experimental data was obtained using
\begin{equation}\label{eq:tau}
  \tau(V)=\tau_0 e^{-V/V_0},
\end{equation}
where $\tau_0 $ and $V_0$ are fitting parameters, see Fig. \ref{fig:0}(d).  Eq.~(\ref{eq:tau}) indicates that the resistance switching in ECM cells is an activated process (an energy barrier must be overcome to change the cell resistance). We note that the exponent in Eq.~(\ref{eq:Poisson}) is the occupation probability of the off-state, while $\Delta t / \tau$ is the probability of switching within the time interval $\Delta t$ given that the cell is in the off-state.

In what follows we assume that the on-to-off switching is also described by a Poisson distribution, albeit parameterized differently. Specifically, the switching rates (inverses of the switching times) are calculated as
 \begin{eqnarray}
 \gamma_{0\rightarrow 1}(V)=\left\{ \begin{array}{cl}
\left( \tau_0 e^{-V/V_0}\right)^{-1},& V>0 \\
0 & \textnormal{otherwise}
\end{array}\right. \; , \label{eq:gamma01}\\
 \gamma_{1\rightarrow 0}(V)=\left\{ \begin{array}{cl}
\left( \tau_1 e^{-|V|/V_1}\right)^{-1},& V<0 \\
0 & \textnormal{otherwise}
\end{array}\right. \; . \label{eq:gamma10}
 \end{eqnarray}

\begin{figure}[tb]
  \centering
   (a) \includegraphics[width=0.75\columnwidth]{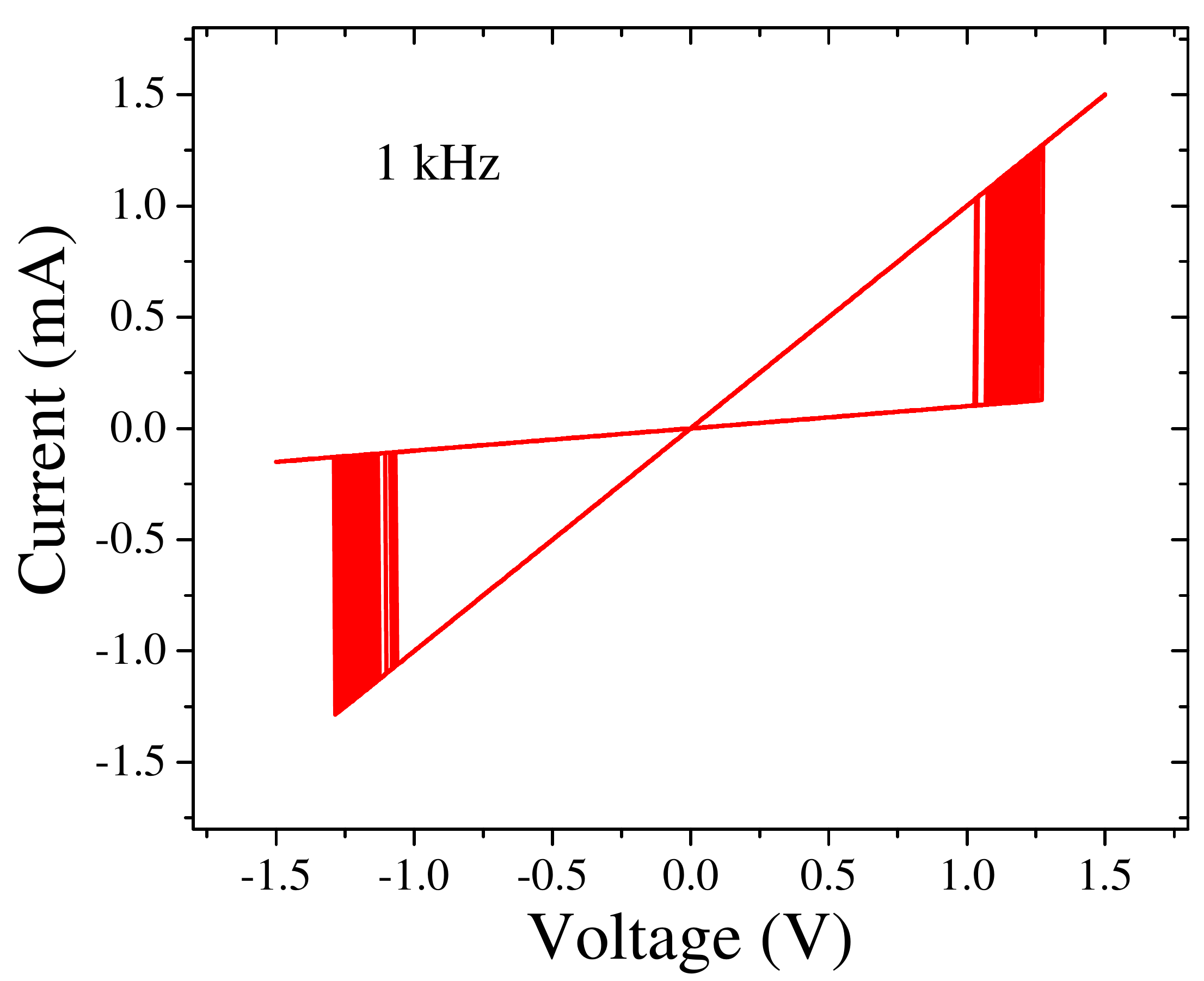} \\
   (b) \includegraphics[width=0.75\columnwidth]{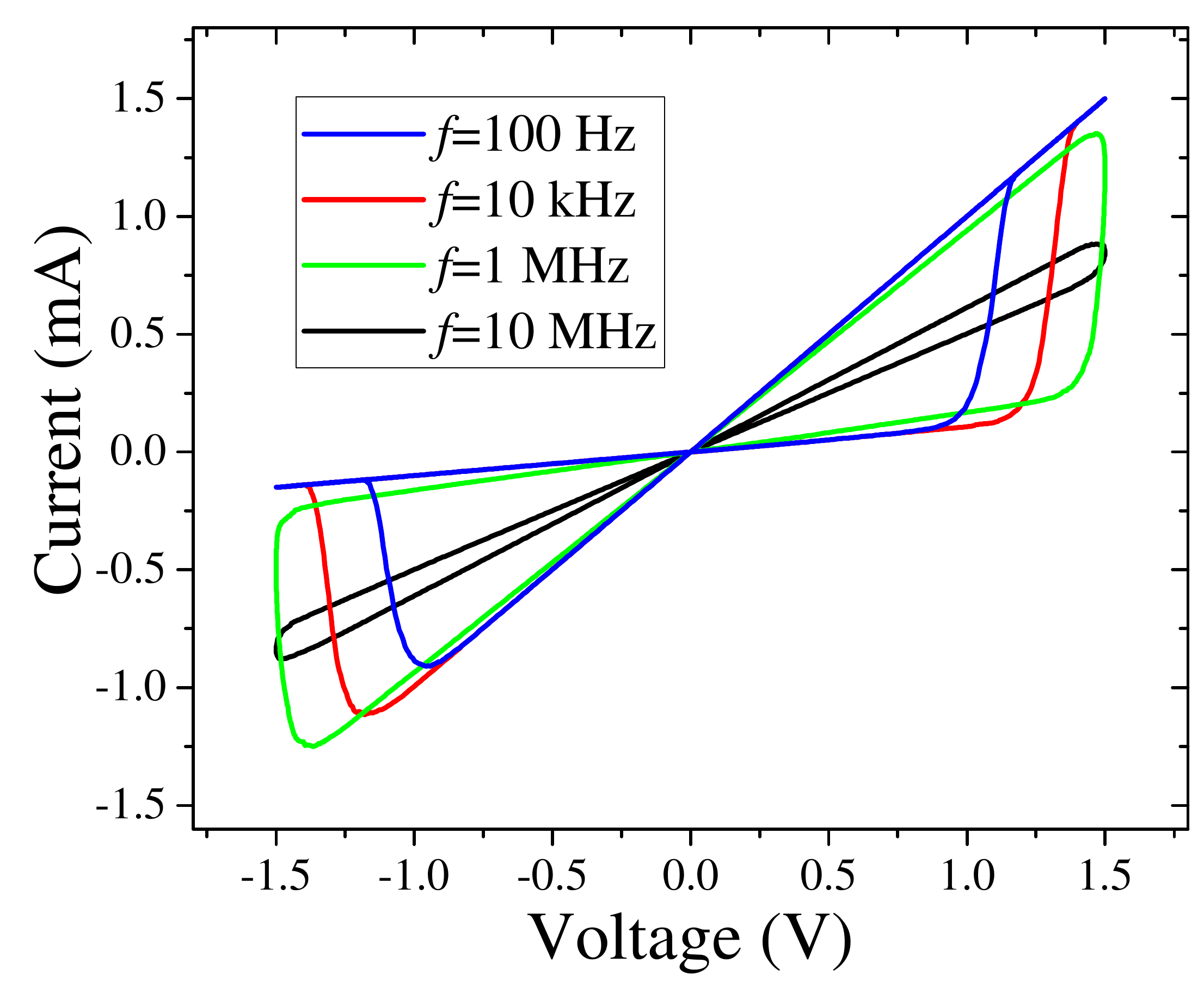}
  \caption{(a) $I-V$ curve of a probabilistic binary memristor. This plot was obtained using the parameter values $\tau_0=\tau_1=3 \cdot 10^5$~s, $V_0=V_1=0.05$~V, $R_{on}=1$~k$\Omega$, $R_{off}=10$~k$\Omega$ and plotted for 100 cycles of $1.5$~V amplitude $1$~kHz frequency sinusoidal voltage. (b) Averaged $I-V$ curves (over a 1000 periods of sinusoidal voltage) plotted for several applied voltage frequencies.} \label{fig:1}
\end{figure}

To graphically represent Eqs.~(\ref{eq:gamma01})-(\ref{eq:gamma10}),  Fig.~\ref{fig:1} shows the current-voltage curves of a probabilistic binary memristor.
In particular, Fig.~\ref{fig:1}(a) demonstrates a very stochastic behavior in the switching region, with a  variability from cycle to cycle. After the averaging (Fig.~\ref{fig:1}(b)), the current-voltage curves resemble the
curves in deterministic models. We emphasize that in  Fig.~\ref{fig:1}(b) the hysteresis collapses at high frequencies -- a well-known feature of the deterministic memristive behavior~\cite{chua76a}.
The collapse is explained by the fact that at high frequencies the duration of the positive/negative half-period is not sufficient for memristor to switch into the on/off-state with probability 1.

\subsection{Numerical simulations}  \label{sec:numerics}

In the simulations presented below, a circuit of $N=10$ memristors initially in the off-state ($R(t=0)=R_{off}$) is considered. It is assumed that the positive applied voltage drives all the memristors into the on-state. In some of our calculations, it is assumed that the memristors are identical with Fig.~\ref{fig:1} parameters. Moreover, the impact of device variability was investigated numerically, assuming a uniform distribution of the parameters $\tau_0$ and $V_0$.

To simulate in-parallel connected memristors (Fig. \ref{fig:2}(a)), each memrisor is subjected to a voltage $V_i=V_a$. A probability for any memristor to switch from the off- to on-state is then generated according to Eq. (\ref{eq:gamma01}) as $\Delta t \gamma_{0\rightarrow 1}(V_a)$, where $\Delta t$ is the simulation time step. This probability is then compared to a random number between zero and one. If the probability is greater than the number generated for that memristor, it switches on. Time is then incremented and the process continues until all memristors are in the on state. The time it takes for the last memristor to switch is then recorded.

To simulate in-series connected memristors (Fig. \ref{fig:2}(b)), a chain of $N$ memristors is subjected to a voltage $V_a$.
The simulation process is the same as in the case of in-parallel memristors, except the voltage across memristors change as the switching progresses. Therefore, at each step, the applied voltage and therefore the switching probabilities are generated for each memristor. As before, the time is then recorded when the final memristor has switched to the on state. For the purpose of comparison, the average voltage across each memristor in the in-parallel and in-series calculations was the same.

This same analysis is also performed for nonidentical memristors. That is $\tau_0$ and $V_0$ are no longer held constant, but are randomly generated for each memristor using uniform distributions.

\subsection{Simplest master equation}

To facilitate the understanding of the master equation approach, let us derive the simplest master equation. The master equation describes how a system composed of probabilistic states evolves. It's a well-known equation in statistical systems~\cite{vanCampen} which is often used to model the time evolution of stochastic processes such as chemical reactions, or diffusion processes, for example.

Consider a single probabilistic memristor connected to a constant voltage source and experiencing off-to-on switching.
The state of the memristor can be represented by the probabilities of finding it in the off- and on-state, $p_0(t)$ and $p_1(t)$, respectively. Clearly, these probabilities change with time as it is more likely that the memristor is found in the on-state as time evolves. If the initial state of memristor is off, then $p_0(t=0)=1$ and $p_1(t=0)=0$. Moreover, since the memristor can be found definitely in the on- or off-state with a unit probability (no other states available), $p_0(t)+p_1(t)=1$.

Next, the probability of switching during the time interval $t$ to $t+\Delta t$ is given by the product of the probability that the memristor is still in the off-state, $p_0(t)$, and the switching probability
$\Delta t / \tau(V)$. Therefore, the occupation probability of the on-state changes as
$$
p_0(t+\Delta t)=p_0(t)-p_0(t)\frac{\Delta t}{\tau(V)}
$$
or
$$
\frac{p_0(t+\Delta t)- p_0(t)}{\Delta t}=-\frac {p_0(t)}{\tau}.
$$
In the limit of $\Delta t \rightarrow 0$, and using $p_0(t)+p_1(t)=1$ we obtain
$$
\frac{\textnormal{d}p_0(t)}{\textnormal{d}t}=-\frac {p_0(t)}{\tau(V)}, \;\;\;\; \frac{\textnormal{d}p_1(t)}{\textnormal{d}t}=\frac {p_0(t)}{\tau(V)}
$$
that is the simplest master equation. It is not difficult to find the solutions of the above equations and verify that the probability of switching given by Eq.~(\ref{eq:Poisson}) (divided by $\Delta t$)
enters into their right-hand sides. Therefore, the change in the probability of finding the memristor in a certain state during some time interval is given by the probability of switching during that same time interval with the appropriate sign.

\section{Master equation} \label{sec:3}

\subsection{General framework}

\begin{figure}[tb]
  \centering
     \includegraphics[width=0.85\columnwidth]{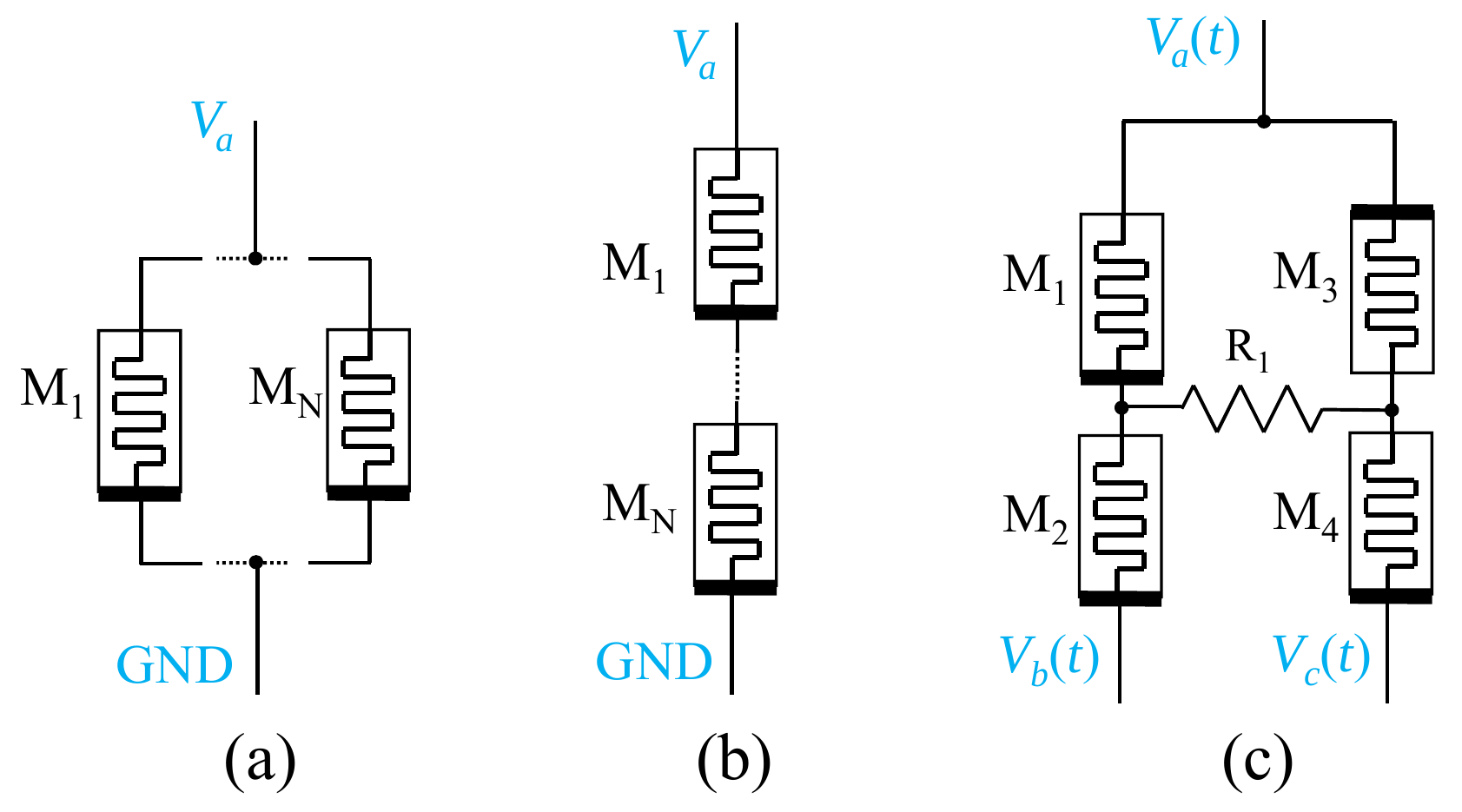}
  \caption{Memristive networks considered in this paper: (a) $N$ memristors connected in-parallel, (b) $N$ memristors connected in-series, and (c) circuit combining memristors, resistors, and subjected to several voltage sources.}\label{fig:2}
\end{figure}

Consider a network composed of $N$ probabilistic memristors, some (or no) resistors, voltage and/or current sources (for an example see Fig. \ref{fig:2}(c)). There are $2^N$ possible network states corresponding to various combinations of the memristor states.
Let's use $\Theta=(...kji)$ to denote a particular network state. Here, $i$ is the state of the first memristor (0/1 for the off/on-state), $j$ is the state of the second memristor, and so on. For a particular network state $\Theta$, the voltage across $m$-th memristor, $V^m_\Theta$, can be found using Kirchhoff's circuit laws~\footnote{Note that the sign of $V^m_\Theta$ depends on the memristor connection polarity.}.

Generally, each realization of circuit dynamics is unique as the time moments when the switchings occur cannot be predicted deterministically. Starting from the same initial state and repeating the experiment many times, one can find time-dependent occupation probabilities of network states, $p_{\Theta}(t)$, that describe the circuit evolution on average. These probabilities can be calculated using the master equation.

\begin{figure}[tb]
  \centering
     \includegraphics[height=0.34\columnwidth]{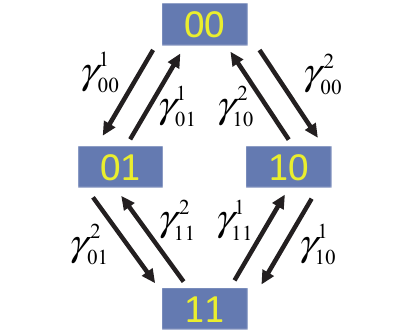}
  \caption{ Full transition scheme for 2 memristors.} \label{fig:X}
\end{figure}

The master equation can be generally written as
\begin{equation}
\frac{\textnormal{d}p_{\Theta}(t)}{\textnormal{d}t}=\sum\limits_{m=1}^{N}\left(\gamma_{\Theta_m}^mp_{\Theta_m}(t)-\gamma_\Theta^m p_{\Theta}(t) \right) \;,
\label{eq:kin}
\end{equation}
where $\Theta_m$ is the network state obtained from $\Theta$ by flipping the state of $m$-th memristor, $\gamma_\Theta^m$ are the transition rates for $m$-th memristor in the configuration $\Theta$ (given by, e.g., Eqs.~(\ref{eq:gamma01}) and (\ref{eq:gamma10})), and $\gamma_{\Theta_m}^m$ is defined similarly~\footnote{Eq.~(\ref{eq:kin}) can be easily extended to the case of multi-state memristors~\cite{Dowling2020}. In
multi-state memristors, the switching between the boundary states occurs through a set of fixed or floating intermediate resistance states involved in the probabilistic process of filament growth or annihilation.}.
We note that the general form of the master equation does {\it not} depend on the circuit topology,
presence or absence of resistors in the circuit, and how the external signals are applied. This information is contained in the voltage-dependent transition rates, $\gamma_\Theta^m$ and $\gamma_{\Theta_m}^m$, that should be evaluated for each network state with the use of Kirchhoff's laws. The full transition scheme for the case of 2 memristors is presented in Fig.~\ref{fig:X}, while
examples of reduced transition schemes are shown in Fig.~\ref{fig:3}. We emphasize that the right-hand side of Eq.~(\ref{eq:kin}) contains only the transitions associated with the flipping the state of a single memristor as, generally, the probability of simultaneous switching is negligibly small (for the theory of kinetic processes it is common to neglect the simultaneous transitions).

The solution of Eq.~(\ref{eq:kin}) can be employed to find various distributions and circuit characteristics on average.
For instance, the average resistance of memristor 1 can be found using
\begin{equation}\label{eq:kin:1}
  \langle R_1(t) \rangle = R_{off} p_{0}^{1}(t)+R_{on} p_{1}^{1}(t),
\end{equation}
where $p_{0}^{1}=\sum\limits_{k,j,...=0,1} p_{...kj0}(t)$, and $p_{1}^{1}=\sum\limits_{k,j,...=0,1} p_{...kj1}(t)$.
Here, the sums are taken over all possible states with a fixed state of memristor 1.
Moreover, various terms in the right-hand side of Eq.~(\ref{eq:kin}) can be of great help in various calculations,
including the calculations of average switching times and their distributions (presented in the next Sec.~\ref{sec:2:case}).

\subsection{Two memristors connected in-series: A case study} \label{sec:2:case}

\begin{figure}[tb]
  \centering
     (a)\includegraphics[height=0.34\columnwidth]{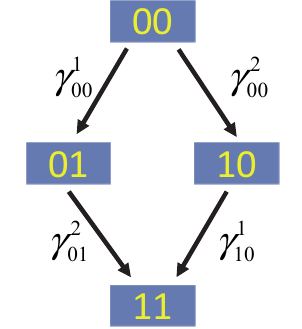}
     (b)\includegraphics[height=0.34\columnwidth]{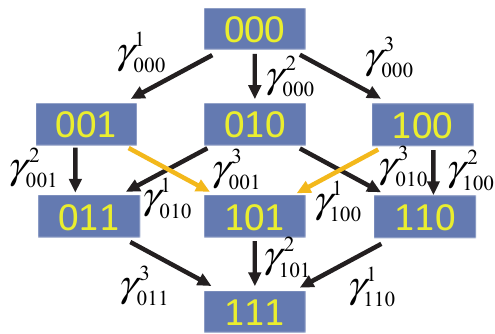}
  \caption{ Reduced transition schemes for (a) 2 and (b) 3 memristors connected in-series or in-parallel, and experiencing the off-to-on switching.} \label{fig:3}
\end{figure}

To exemplify the approach in Eq.~(\ref{eq:kin}), consider a relatively simple yet interesting problem of the resistance switching in
a circuit of two probabilistic binary memristors connected in-series. It is assumed that the memristors are connected to a constant positive voltage, and experience switching from the off- into the on-state. Thus the initial conditions are $p_{00}=1$ and $p_{ij}=0$ for $(i,j)\neq(0,0)$. Since a positive voltage is applied to each memristor, the on- to off- transitions are impossible in the given circuit configuration, therefore, the corresponding switching rates (such as $\gamma_{01}^1$ and $\gamma_{10}^2$) are equal to 0.
Fig.~\ref{fig:3}(a) presents a reduced transition scheme for the problem that includes only the processes occurring at $V_a>0$.

Assuming that the memristors are identical, we set $\gamma_{00}^1=\gamma_{00}^2$,
$\gamma_{01}^2=\gamma_{10}^1$, $p_{01}(t)=p_{10}(t)$. Then Eq.~(\ref{eq:kin}) takes the form
\begin{eqnarray}
  \frac{\textnormal{d}p_{00}(t)}{\textnormal{d}t} &=& -2\gamma_{00}^1p_{00}, \label{eq:twoM:1} \\
  \frac{\textnormal{d}p_{01}(t)}{\textnormal{d}t} &=& \gamma_{00}^1p_{00}-\gamma_{01}^2p_{01}, \label{eq:twoM:2} \\
  \frac{\textnormal{d}p_{11}(t)}{\textnormal{d}t} &=& 2\gamma_{01}^2p_{01}, \label{eq:twoM:3}
\end{eqnarray}
where $\gamma_{00}^1=\gamma_{0\rightarrow 1}(V^1_{00})$, and $\gamma_{01}^2=\gamma_{0\rightarrow 1}(V^2_{01})$. Here,
$V^1_{00}=V_a/2$ is the voltage across memristor 1 in the network state $(00)$, while $V^2_{01}=R_{off}/(R_{on}+R_{off})V_a$ is the voltage across memristor 2 in the network state $(01)$, and $V_a$ is the applied voltage. The solution of Eqs.~(\ref{eq:twoM:1})-(\ref{eq:twoM:3}) reads
\begin{eqnarray}
  p_{00}(t) &=& e^{-2\gamma_{00}^1 t}, \label{sol:1} \\
  p_{01}(t)=p_{10}(t) &=& \frac{\gamma_{00}^1}{\gamma_{01}^2-2\gamma_{00}^1}\left(e^{-2\gamma_{00}^1 t}-e^{-\gamma_{01}^2 t} \right) ,\;\;\\
  p_{11}(t) &=& 1-p_{00}(t)-2p_{01}(t) \label{sol:3}.
\end{eqnarray}

{\it Average network switching time}-- The network switching time is associated with the transition to the
state 11. The switching probability distribution as a function of time is given by the right-hand side of Eq.~(\ref{eq:twoM:3}), $2\gamma_{01}^2p_{01}(t)$.
It can be used to calculate the average switching time according to
\begin{equation}\label{eq:T2}
\langle  T_{11}\rangle =\int\limits_0^\infty t 2\gamma_{01}^2p_{01}(t) \textnormal{d}t =\frac{1}{2\gamma_{00}^1}+\frac{1}{\gamma_{01}^2}.
\end{equation}

The variance $\langle\left( t- \langle T_{11}\rangle\right)^2\rangle$ represents a measure of the cycle-to-cycle variability. To find the variance, the outer averaging is performed as in the above Eq.~(\ref{eq:T2}), with $\langle  T_{11}\rangle$
represented by the right-hand side of Eq.~(\ref{eq:T2}). We find
\begin{eqnarray*}
\langle(t-\langle T_{11}\rangle)^2\rangle&=&2\gamma_{01}^2 \int\limits_0^\infty \left(t-\left[
\frac{1}{2\gamma_{00}^1}+\frac{1}{\gamma_{01}^2}   \right] \right)^2p_{01}(t)\textnormal{d}t  \\
&=&\frac{1}{4(\gamma_{00}^1)^2}+\frac{1}{(\gamma_{01}^2)^2}.
\end{eqnarray*}

{\it Average switching time of memristor 1.}-- This switching time is associated with transitions $00\rightarrow 01$ and
$10\rightarrow 11$. For these processes, the switching probability distribution can be expressed as
\begin{equation}\label{eq:Phia}
  \Phi_{1}(t)=\gamma_{00}^1p_{00}(t)+\gamma_{10}^1p_{10}(t).
\end{equation}
Using Eq.~(\ref{eq:Phia}), one can find
\begin{equation}\label{eq:T2:a}
\langle  T_{1}\rangle =\int\limits_0^\infty t \Phi_{1}(t) \textnormal{d}t =\frac{1}{2\gamma_{00}^1}+\frac{1}{2\gamma_{01}^2}.
\end{equation}

{\it Average resistance of memristor 1.}-- This quantity can be directly calculated using the probabilities (\ref{sol:1})-(\ref{sol:3}) as
\begin{equation}\label{eq:R1}
\langle R_{1}(t) \rangle =R_{off} \left(  p_{00}(t)+ p_{10}(t) \right)+R_{on} \left(  p_{01}(t)+ p_{11}(t) \right).
\end{equation}

It is interesting to compare the switching time of memristors connected in series with the switching time for memristors connected in parallel. The latter is derived in the Appendix~\ref{app:2} (Eq.~(\ref{eq:ap2:5}) for $N=2$). Using
$R_{on}=1$~k$\Omega$, $R_{off}=10$~k$\Omega$, $\tau_0=3\cdot 10^5$~s, $V_0=0.05$~V, $V_a=2$~V (in-series), and  $V_a=1$~V (in-parallel), we find
{\begin{eqnarray}
 \langle  T_{11} \rangle&=& 309\;\mu\textnormal{s}, \\
  \langle  T_{\|,2} \rangle  &=& 928\;\mu\textnormal{s}.
 \end{eqnarray}}
This estimation indicates that the switching of memristors connected in-series occurs significantly faster compared to the switching of in-parallel connected ones. Physically, such behavior can be explained by the voltage divider effect where the switching of one memristor leads to a voltage increase across another accelerating its switching.

\subsection{More complex cases}

Using numerical simulations, we studied the switching in the networks of $N=10$ memristors connected in-series and in-parallel. The simulation approach is described in Sec.~\ref{sec:numerics}. We investigated the networks of identical and non-identical memristors. In the case of identical
memristors, we have verified that numerical results are in agreement with analytical results presented in Appendix~\ref{app:1}. In fact, one of our main analytical findings is the expression for the network switching time, Eq.~(\ref{eq:App:sol13}), which can be rewritten as
\begin{equation}
\label{eq:App:sol13a}
  \left< T_N\right>=\sum_{j=0}^{N-1}\frac{1}{(N-j)\gamma_j},
\end{equation}
where $\gamma_j$ is defined below Eq.~(\ref{eq:App:kin}), and can be evaluated with the help of Eq.~(\ref{eq:gamma01}). We emphasize that Eq.~(\ref{eq:App:sol13a}) also describes the off-to-on transition in the network of in-parallel connected memristors (see Eq. (\ref{eq:ap2:6})), and can be used to model the on-to-off transitions (with a proper selection of switching rates).

\begin{figure}[h]
  \centering   \includegraphics[width=0.75\columnwidth]{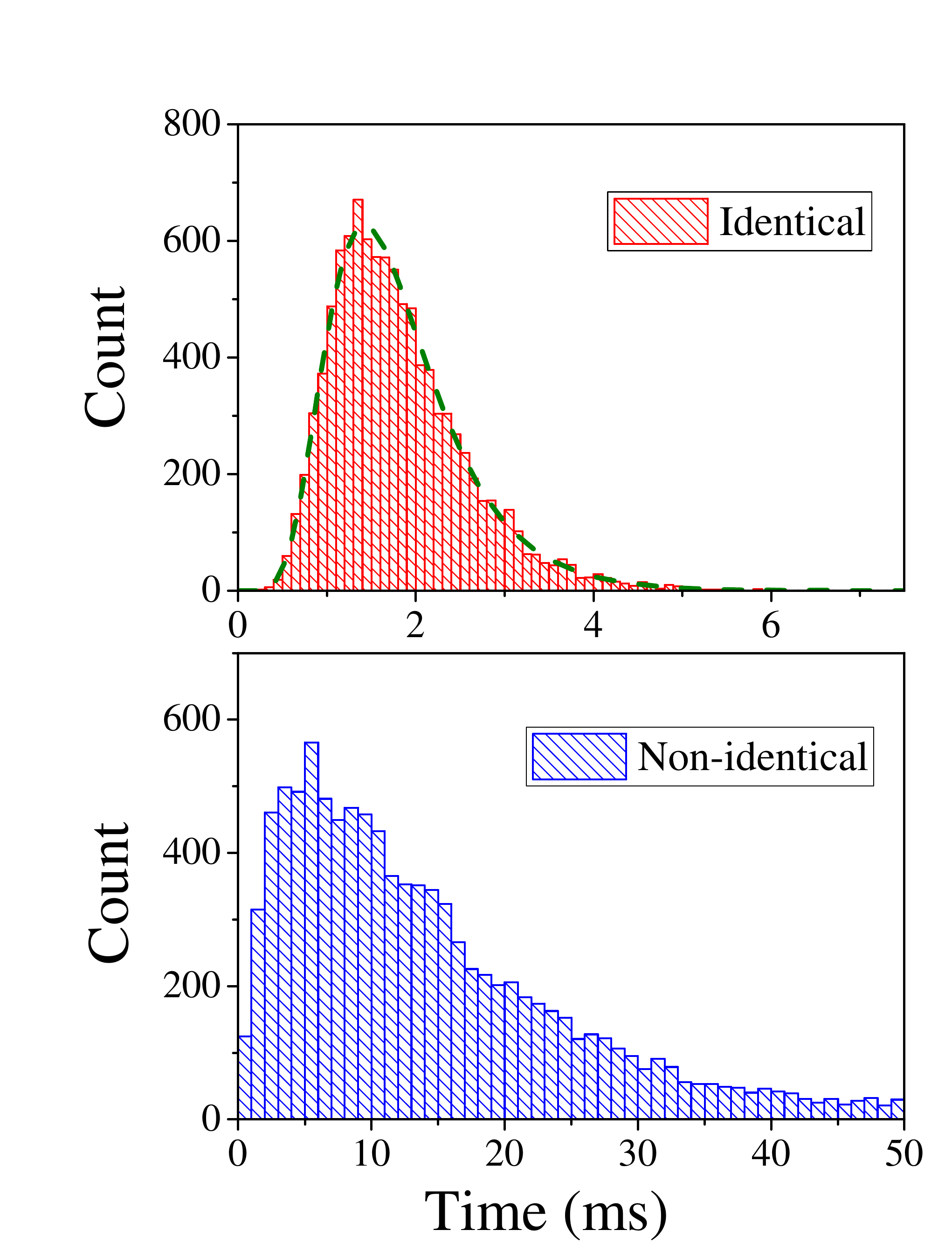}
\caption{ Switching time distributions for network of $N=10$ memristors connected in-parallel switching from the off- to on-state with $V_a=1$~V found in $10^4$ numerical simulations. The identical memristors have constants $\tau_0=3 \cdot 10^5$~s, and $V_0=0.05$~V. The non-identical memristors are characterized by random flat distributions of $\tau_0$ and $V_0$ in the intervals $[2\cdot 10^5,4\cdot 10^5]$~s and
 $[0.04,0.06]$~V, respectively. The mean switching time is $1.81$~ms for identical memristors and $15.3$~ms for non-identical memristors. The bin size is $0.1$~ms (top histogram) and 1~ms (bottom histogram). The dashed line overlaying the top histogram is found analytically using the master equation approach (see text for details).} \label{fig:4}
\end{figure}

Fig.~\ref{fig:4} shows the distributions of switching times in the networks of identical and non-identical memristors connected in-parallel. In the case of non-identical memristors,  $\tau_0$ and $V_0$ are determined by probabilistic distributions for each memristor to see if the randomness of $\tau_0$ and $V_0$ have any significant effect on the network dynamics. For the sake of simplicity, random flat distributions are used. According to Fig.~\ref{fig:4},  the randomness of $\tau_0$ and $V_0$ significantly broadens the distribution of switching times in the case of in-parallel connected memristors.
 As memristors connected in-parallel switch independently, their network switching time depends significantly on the slowest switching memristor, which, statistically, has a longer characteristic switching time than that of identical memristors.

\begin{figure}[h]
  \centering
     \includegraphics[width=0.75\columnwidth]{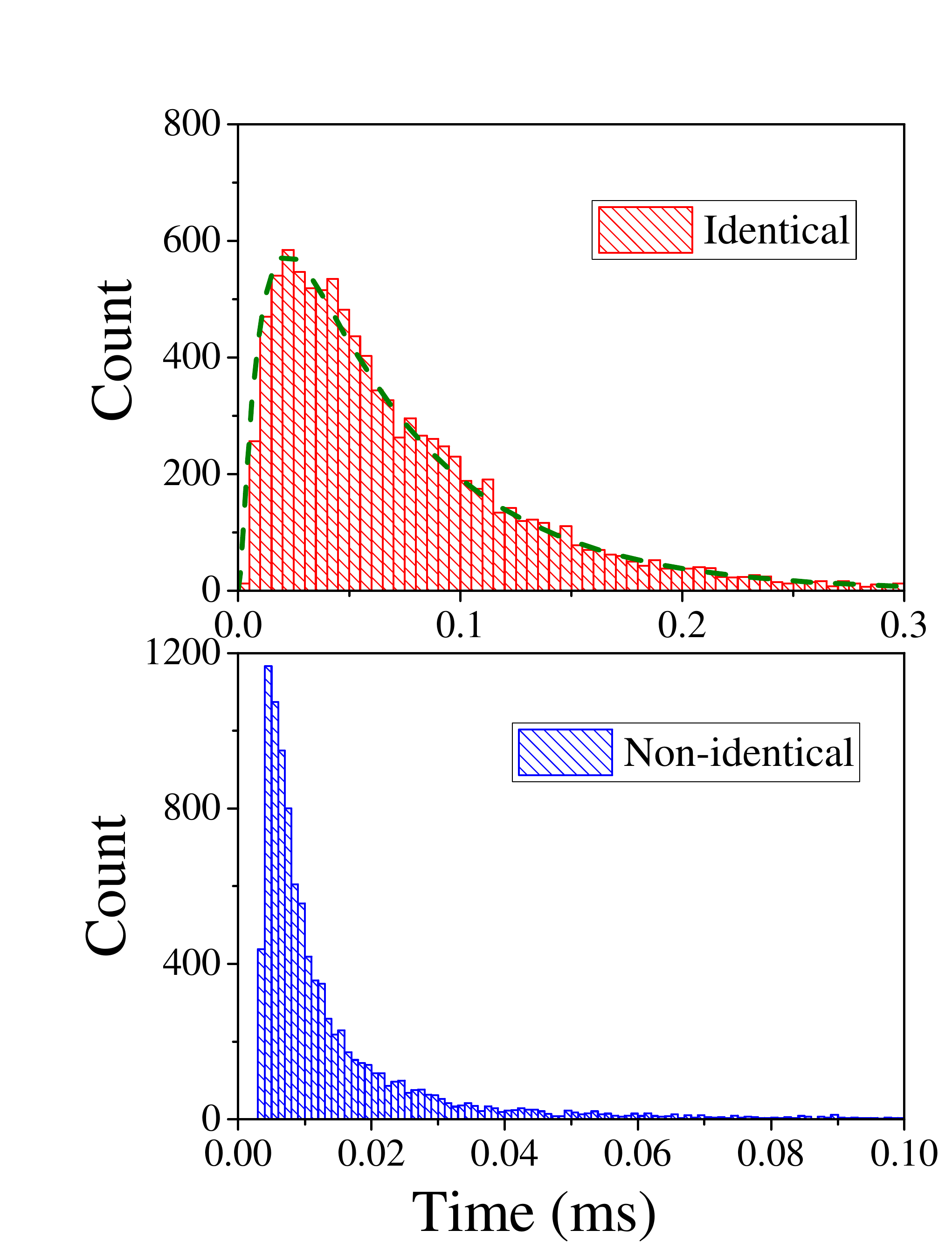}
\caption{Switching time distributions for network of $N=10$ memristors connected in-series switching from the off- to on-state with $R_{off}/R_{on}=10$ at $V_a=10$~V found in $10^4$ numerical simulations. The memristor parameters are as in Fig.~\ref{fig:4}. The mean switching time is $74.2$~$\mu$s for identical and $16.4$~$\mu$s for non-identical memristors. The bin size is $5$~$\mu$s (top histogram) and 1~$\mu$s (bottom histogram).   The dashed line overlaying the top histogram is found analytically using the master equation approach (see text for details).
 } \label{fig:5}
\end{figure}

The distributions of network switching time for identical and non-identical memristors connected in-series are presented in Fig.~\ref{fig:5}.  We note that Figs.~\ref{fig:4} and \ref{fig:5} were obtained assuming the same voltage across each memristor on average. In the case of in-series connected memristors (Fig.~\ref{fig:5}), the voltages across memristors were recalculated at each time step according to the instantaneous network configuration.
Generally, in-series connected memristors switch faster than the memristors connected in-parallel. This is explained by a cascading effect for in-series connected memristors: The switching to the on-state of one generates an increased probability to switch for the remaining off-state memristors. We note that the shorter (on-average) network switching time for the case of non-identical memristors in Fig.~\ref{fig:5} is due to the important role of the fastest switching memristor in the network.

Our numerical results for both in-parallel and in-series connected identical memristors are in perfect agreement with analytical results. The distribution of network switching time is associated with the dynamics of occupation of the last state, and is simply given by $\textnormal{d}p_{11..11}/\textnormal{d}t$. In the case of in-parallel connected memristors, the time derivative of Eq.~(\ref{eq:ap2:3}) results in the switching time distribution
\begin{equation*}
  N\gamma_0^1\left( 1-e^{-\gamma_0^1 t} \right)^{N-1}e^{-\gamma_0^1 t}.
\end{equation*}
This distribution (appropriately normalized) is presented by the dashed curve in Fig.~\ref{fig:4}. The network switching time for in-parallel connected memristors is calculated using Eq.~\ref{eq:ap2:6} expression, and is exactly $1.81$~ms.

In the case of in-series connected memristors, $\textnormal{d}p_{11..11}/\textnormal{d}t$ is calculated using $b_N p_{N-1}(t)$ with the help of Eq.~(\ref{eq:App:sol6}). This distribution is plotted in Fig.~\ref{fig:5} (appropriately normalized). The perfect agreement with the numerical result is evident. The in-series switching time found using Eq.~(\ref{eq:App:sol13a}) is 72.4~$\mu$s, which is slightly shorter than that found in our numerical simulations ($74.2$~$\mu$s).

\section{Correlation functions} \label{sec:4}

\subsection{General approach}

When the memristors interact through a circuit, correlations between their states develop. Correlation functions~\cite{vanCampen} are a common tool used for their description. For instance, for two memristors $i$ and $j$, the two-times correlation function can be defined as
\begin{equation}\label{eq:corr:1}
  K_{ij}(t,s)=\langle  R_i(t)R_j{(t+s)}\rangle - \langle  R_i(t)\rangle\langle R_j{(t+s)}\rangle,
\end{equation}
where $s$ defines a second moment of time, which is shifted from $t$ by $s$. Similarly, we can define  the auto-correlation function
\begin{equation}\label{eq:corr:a}
  K_{ii}(t,s)=\langle  R_i(t)R_i{(t+s)}\rangle - \langle  R_i(t)\rangle\langle R_i{(t+s)}\rangle,
\end{equation}
which allows us to find, in particular, the variance $\mathbf{Var}(R_i)$ of the resistance of selected memristor $i$, by substituting $s=0$ into Eq.~(\ref{eq:corr:a}).

\subsection{Two memristors connected in-series}

To derive correlation functions analytically, we first introduce a joint probability distribution function
\begin{equation}\label{eq:joint_p}
  \Phi (t_1,t_2)\textnormal{d}t_1\textnormal{d}t_2=\gamma_{00}^1 e^{-2\gamma_{00}^1 t_2}\textnormal{d}t_2 \gamma_{01}^2e^{-\gamma_{01}^2 (t_1-t_2)}\textnormal{d}t_1.
\end{equation}
Eq.~(\ref{eq:joint_p}) describes the probability of switchings of memristor 1 in the time interval from $t_1$ to $t_1+\textnormal{d}t_1$, and
memristor 2 in the time interval from $t_2$ to $t_2+\textnormal{d}t_2$ in the assumption of $t_1>t_2$. Here, the term $\gamma_{00}^1 e^{-2\gamma_{00}^1 t_2}\textnormal{d}t_2$ is the probability of memristor 2 independently switching and $\gamma_{01}^2e^{-\gamma_{01}^2 (t_1-t_2)}\textnormal{d}t_1$ is the probability of memristor 1 switching assuming memristor 2 has already switched at $t=t_2$. The case of $t_2>t_1$ is  described by the right-hand side of Eq.~(\ref{eq:joint_p}) with $1\leftrightarrow 2$. We note that in Eq.~(\ref{eq:joint_p}), one can recognize the well-known expression for the conditional probability.

Eq.~(\ref{eq:joint_p}) can be used to re-derive various quantities already discussed in Sec.~\ref{sec:2:case}.
For the convenience of the reader, some of the relevant relations are provided in Appendix~\ref{app:3}.
 To derive the correlation function (\ref{eq:corr:1}) we note that the resistance as a function of time can be presented as
\begin{equation}\label{eq:R(t)}
  R_i(t)=R_{off}+(R_{on}-R_{off})H(t-t_i),
\end{equation}
where $H(..)$ is the Heaviside step function, and $t_i$ is the switching time of the memristor $i$. The average of $R_i(t)$ can be found using Eq.~(\ref{eq:R1}).
The calculation of $K_{12}(t,s)$ in Eq.~(\ref{eq:corr:1}) involves finding the average of the product of Heaviside functions
\begin{eqnarray}
  \nonumber &&\langle H(t-t_1) H(t+s-t_2) \rangle = \\
 \nonumber &=&\int\limits_0^\infty\int\limits_0^\infty\Phi(t_1,t_2) H(t-t_1) H(t+s-t_2)\textnormal{d}t_2\textnormal{d}t_1 = \;\; \\
  &=&p_{10}(t)+p_{11}(t)-e^{-\gamma^2_{01}s}p_{01}(t),
\end{eqnarray}
where we took into account Eqs.~(\ref{eq:F1}), (\ref{eq:Phia}), and (\ref{eq:joint_p}).
By substituting  Eq.~(\ref{eq:R(t)}) into Eq.~(\ref{eq:corr:1}) for the cases $i=1$ and $j=2$ we get
\begin{eqnarray}\label{eq:K2a1_}
  \nonumber \frac{K_{12}(t,s)}{\left( R_{off}-R_{on} \right)^2}&=&\langle H(t-t_1) H(t+s-t_2) \rangle\\
 \nonumber    &-&\langle H(t-t_1)\rangle\langle H(t+s-t_2) \rangle.
\end{eqnarray}
This leads to the following expression for the correlation
function
\begin{equation}\label{eq:K2a1}
  \frac{K_{12}(t,s)}{\left( R_{off}-R_{on} \right)^2}=[1-p_0(t)]p_0(t+s)-p_{01}(t)e^{-\gamma_{01}^2s},
\end{equation}
where $p_0(t)=p_{00}(t)+p_{01}(t)$.

The same technique can be used to calculate the auto-correlation function $K_{ii}(t,s)$ defined by Eq.~(\ref{eq:corr:a}). In this case, it is even simpler to do
it because we need only the switching probability distribution Eq.~(\ref{eq:Phia}). As a result we get for the auto-correlation function
\begin{equation}\label{eq:K2a2}
  \frac{K_{ii}(t,s)}{\left( R_{off}-R_{on} \right)^2}=[1-p_0(t)]p_0(t+s).
\end{equation}

\subsection{More complex cases}

\begin{figure}[tb]
  \centering
     \includegraphics[width=0.75\columnwidth]{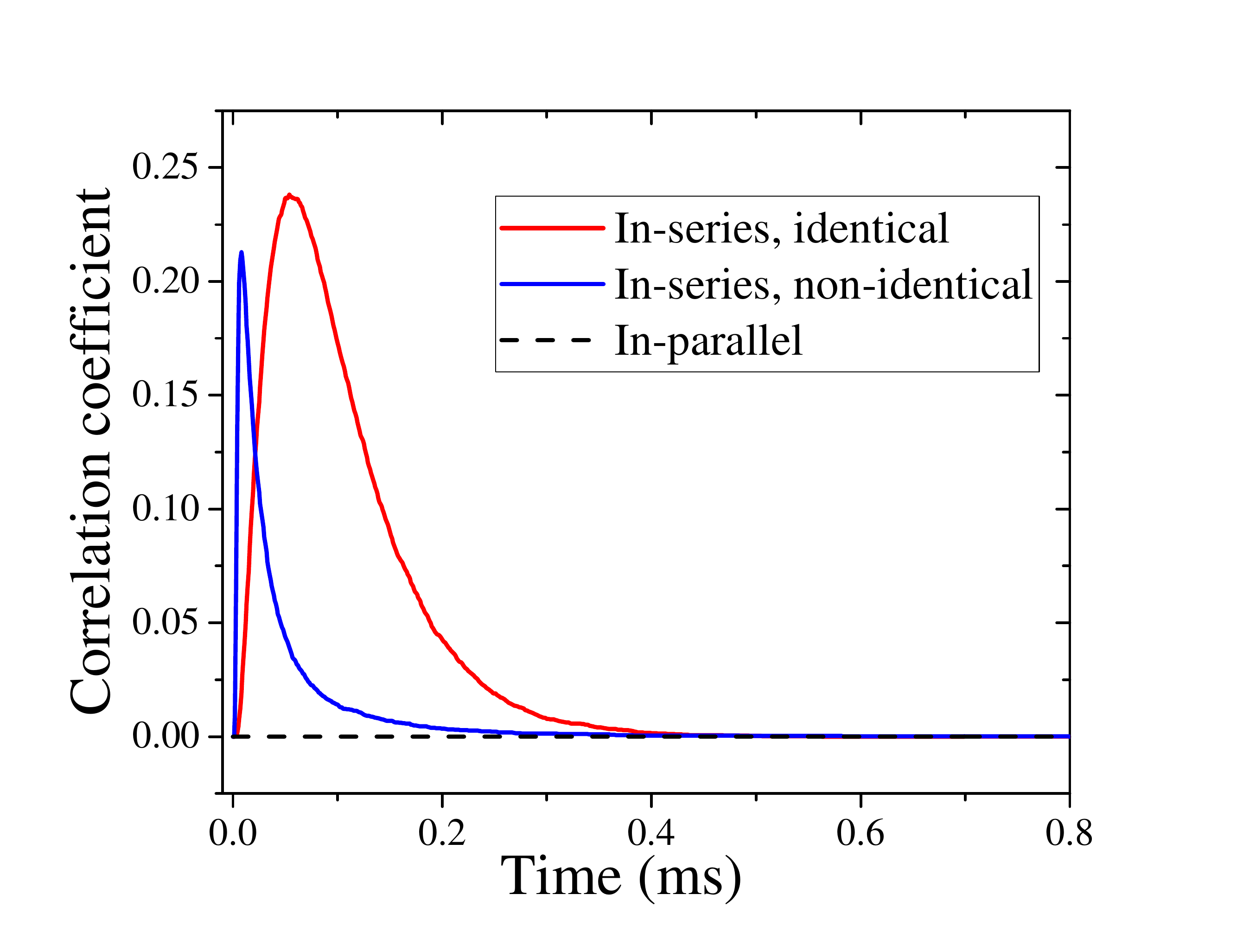} \\
\caption{ One-time correlation function $\widetilde{K}_{i,j}(t)$ for a set of $N=10$ identical and non-identical memristors. The memristor parameters are the same as in Fig.\ref{fig:4}. } \label{fig:6}
\end{figure}

A normalized one-time correlation function for two randomly chosen memristors $i$ and $j$ can be calculated using
\[\widetilde{K}_{ij}(t)\equiv K_{ij}(t,0) =\frac{<R_i(t)R_j(t)>-<R_i(t)><R_j(t)>}{(R_{off}-R_{on})^2},
\]
where $R_i(t)$ is the resistance of memristor $i$ at time $t$. The above expression was evaluated numerically for $N=10$ memristive networks. The results are shown in Figure \ref{fig:6} for the networks of identical and non-identical memristors.

Several features in Fig.~\ref{fig:6} can be mentioned here. First, at the initial moment of time $\widetilde{K}_{i,j}(0)=0$ as the initial state of network is deterministic (all memristors are in the off-state initially). Second, the in-series correlation functions have a maximum when the probabilities of $R_{on}$ and $R_{off}$ are approximately the same. Moreover, the maximum value of these functions does not exceed 0.25. Third, at long times, the in-series functions approach zero as the memristor states at long times are nearly deterministic (all memristors end up in the on-state). Finally, $\widetilde{K}_{i,j}(t)$ for in-parallel connected memristors is always zero as such memristors do not interact through the network. Therefore, correlations among them do not develop.

\section{Discussion and Conclusion} \label{sec:5}

The modeling of probabilistic memristive networks presents opportunities and challenges. The opportunities open up as there is an increasing interest in the stochastic computing~\cite{gaba2014memristive,7827976,Alahmadi17a,Ielmini2018,carboni2019stochastic,hirtzlin2019stochastic} and neuromorphic computing with stochastic synapses~\cite{Neftci}, and, in principle, all ReRAM devices exhibit a certain level of  stochasticity. The fact that the probabilistic memristive networks can be described in terms of the master equation offers strong possibilities to simulate various processes ranging from chemical reactions to radioactive decay in hardware. The challenges are due to the complexity of probabilistic modeling. In the case of binary memristors, the number of network states increases as $2^N$. Therefore, to describe even modest networks, say, of $N=20$ memristors, already more than $10^6$ network states are required~\footnote{In the case of symmetries some simplifications are possible (e.g., identical memristors, etc.).}.

SPICE simulations of probabilistic memristive networks can be performed similarly to the SPICE modeling of chemical reactions~\cite{madec2017modeling,8585653}. For this purpose, the master equation (\ref{eq:kin}) can be mapped to an electronic circuit with the capacitor charge representing the state occupation probabilities $p_\Theta(t)$, and other components such as voltage-controlled current sources used to represent the right-hand side of Eq. (\ref{eq:kin}). Depending on the a-priori knowledge of driving conditions, either a full transition scheme (such as in Fig.~\ref{fig:X}) or partial one (such as in Fig.~\ref{fig:3}) can be implemented. Details of SPICE modeling of probabilistic memristive networks can be found in our recent preprint posted on arxiv~\cite{dowling2020modeling}.

Electrochemical metallization cells have been considered as binary memristors~\cite{suri2013bio,Truong2014}, and currently they are the most suitable type of ReRAM cells to test our theory. In fact, the model parameters used in this work (listed in Fig.~\ref{fig:1} caption) were extracted from a fitting curve in Ref.~\cite{jo2009programmable} with a subsequent scaling of $V_0$ in the assumption of $\sim 20$~nm a-Si layer. However, the extracted value of $\tau_0=3\cdot 10^5$~s is quite short~\footnote{This constant is a measure of the information storage time at zero applied voltage.}. A more realistic (in terms of the long-time information storage capability) model -- an adaptive probabilistic threshold model (APTM) -- is formulated in Appendix~\ref{app:4}. The hysteretic curve of the APTM model are qualitatively similar to Fig. \ref{fig:1} in the main text. In certain cases, the switching in VCM cells can also be considered as binary~\cite{Fleck16a,Cueppers19a}.

Finally, we note that care must be taken when the binary model is used to simulate experiments with physical devices.
A limitation is related to the fact that in electrochemical metallization cells the off-to-on transition may occur in a step-by-step fashion when the filament advances through several hopping sites~\cite{jo2009programmable}. Moreover, in the resistor-ECM cell circuits the filament growth may be reduced due to the voltage divider effect~\cite{jo2009programmable}.
In principle, the approach presented in this paper can be generalized to more complex circuits involving also capacitors and/or inductors. However, the description of such circuits becomes more complicated as additional continuous variables (such as the ones representing the capacitor charge) need to be taken into account. We plan to explore this direction in the future work.


To conclude, the modeling of stochastic memristors and their circuits is still in a nascent stage compared to the case of deterministic devices. In this paper we have introduced a master equation-based approach to model networks of probabilistic memristors. This approach provides very detailed information about the system including various switching times, occupation probabilities, and correlation functions. This work advances the field of memristor circuits~\cite{chua2019handbook,9050655} by introducing the methodology to model networks of probabilistic memristors, and by finding the solution of a master equation in several model cases.  We expect that the suggested approach will find a wide range of applications, including small, intermediate, and large~\cite{Noh11a,C6CP05187A} probabilistic networks.
\appendix
\numberwithin{equation}{section}
\numberwithin{figure}{section}

\section{Switching of $N$ memristors connected in-series}   \label{app:1}

\setcounter{equation}{0}
\renewcommand\theequation{A.\arabic{equation}}

Consider the dynamics of $N$ identical probabilistic memristors connected in-series to a constant voltage source $V_a$.
It is assumed that at $t=0$ all the memristors are in the off-state, and the applied voltage induces their switching into the on-state.

\subsection{Equations}
We simplify the kinetic equation~(\ref{eq:kin}), made possible due to symmetric initial conditions and similarity of memristors. In this situation the probabilities of all network states with the same number of memristors in the on-state are the same (for instance, for $N=2$, $p_{01}(t)=p_{10}(t)$). To simplify the notation, in this Appendix we use $p_m$ to denote the probability of a state with $m$ memristors in the on-state. Then, Eq. (\ref{eq:kin}) can be rewritten in the form
\begin{equation}\label{eq:App:kin}
  \begin{split}
    \frac{\textnormal{d}p_0}{\textnormal{d}t}&=- N \gamma_0p_0,  \\
   \frac{\textnormal{d}p_m}{\textnormal{d}t}&=m\gamma_{m-1}p_{m-1}-(N-m)\gamma_mp_m,
  \end{split}
\end{equation}
where $m$ changes from 1 to $N$, $\gamma_{m-1}$ is the transition rate from $p_{m-1}$ to $p_{m}$, and $\gamma_m$ is defined similarly.

We note that the occupation probabilities are subjected to the constraint
\begin{equation}\label{eq:App:constr}
  \sum\limits_{m=0}^{N} \begin{pmatrix}N\\m\end{pmatrix} p_m(t)=1.
\end{equation}
Here, the binomial coefficients $\begin{pmatrix}N\\m\end{pmatrix}$ take into account the number of states with the same number of memristors in the on-state. Differentiating Eq. (\ref{eq:App:constr}) with respect to
$t$ we get
\begin{equation}\label{eq:App:constr1}
  \sum\limits_{m=0}^{N} \begin{pmatrix}N\\m\end{pmatrix} \frac{\textnormal{d}p_m}{\textnormal{d}t}=0.
\end{equation}

In what follows, Eq.~(\ref{eq:App:kin}) is solved analytically using the following initial conditions: $p_0(t=0)=1$, $p_{i}(t=0)=0$ for $i=1,..,N$.

\subsection{Building solution}
Defining $a_m=(N-m)\gamma_m$ and $b_m=m\gamma_{m-1}$ Eqs.~(\ref{eq:App:kin}) can be rewritten as
\begin{equation}\label{eq:App:kin1}
  \begin{split}
    \frac{\textnormal{d}p_0}{\textnormal{d}t}&=- a_0p_0,  \\
   \frac{\textnormal{d}p_m}{\textnormal{d}t}&=b_mp_{m-1}-a_mp_m,
  \end{split}
\end{equation}
For the first of the above equations, the solution is
\begin{equation}
p_0(t)=e^{-a_0t}.
\end{equation}
For $m=1$, the equation is
\begin{equation}
\nonumber \frac{\textnormal{d}p_1}{\textnormal{d}t}=b_1p_0-a_1p_1,
\end{equation}
whose solution can be presented as
\begin{equation}
p_1(t)=b_1\left(\frac{e^{-a_0t}}{a_1-a_0}+\frac{e^{-a_1t}}{a_0-a_1}\right).
\end{equation}
Finally, consider the case of $m=2$. The solution of
$$\frac{\textnormal{d}p_2}{\textnormal{d}t}=b_2p_1-a_2p_2$$
is given by
\begin{eqnarray}
p_2=b_1b_2\left[\frac{e^{-a_0t}}{(a_1-a_0)(a_2-a_0)}+\frac{e^{-a_1t}}{(a_0-a_1)(a_2-a_1)}+\right. \nonumber \\
\left.+\frac{e^{-a_2t}}{(a_0-a_2)(a_1-a_2)}\right]. \;\; \label{eq:abx}
\end{eqnarray}

\subsection{Solution}

Based on the above analysis, the probability $p_m(t)$ involves $m$ exponentially decaying terms. Therefore, at step $m$ we can write
\begin{equation}\label{eq:App:sol}
p_m(t)=\sum_{i=0}^{m}C_i^me^{-a_it}
\end{equation}
and at step $m-1$
\begin{equation}\label{eq:App:sol1}
p_{m-1}(t)=\sum_{i=0}^{m-1}C_i^{m-1}e^{-a_it}.
\end{equation}
Here, $C_i^m$ is the $i$-th pre-exponential factor at the step $m$, and $C_i^{m-1}$ is the one at the step $m-1$.

To find the relation among the coefficients $C_i^j$ at different steps, consider Eq.~(\ref{eq:App:kin1}).
Substituting $p_{m-1}$ in the form of Eq.~(\ref{eq:App:sol1}) into Eq.~(\ref{eq:App:kin1}), one can find
\begin{equation}
 \nonumber \frac{\textnormal{d}p_m}{\textnormal{d}t}+a_mp_m = b_m\sum_{i=0}^{m-1}C_i^{m-1}e^{-a_it}\\
\end{equation}
Multiplying both sides by the integrating factor $e^{a_mt}$ leads to
\begin{eqnarray} \label{eq:App:sol4}
 \nonumber  \frac{\textnormal{d}}{\textnormal{d}t}\left(p_me^{a_mt}\right)=b_m\sum_{i=0}^{m-1}C_i^{m-1}e^{a_mt-a_it} \;\; , \\
   p_m(t)=\sum_{i=0}^{m-1}\frac{b_m}{a_m-a_i}C_i^{m-1}e^{-a_it}+C_m^me^{-a_mt}.
\end{eqnarray}
Here, $C_m^m$ is the integration constant, that can be determined from the initial condition $p_m(t=0)=0$. Importantly,
Eq.~(\ref{eq:App:sol4}) shows explicitly how the pre-exponential factors $C_i^j$ evolve from step to step: $C_i^m=b_m/(a_i-a_m)C_i^{m-1}$, $i<m$.

As we prove below,
$C_m^m$ can be presented as
\begin{equation}\label{eq:App:sol5}
  C_{m}^{m}=\prod_{i=0}^{m-1}\frac{b_{i+1}}{a_i-a_m}.
\end{equation}
Therefore, the occupation probability of a state with $m$ memristors in the on-state is given by
\begin{equation}
\nonumber  p_m(t)=\frac{b_m}{a_m-a_0}C_0^{m-1}e^{-a_0t}+...+ \\
\end{equation}
\begin{equation}
\nonumber \frac{b_m}{a_m-a_{m-1}}C_{m-1}^{m-1}e^{-a_{m-1}t} +\\
\end{equation}
\begin{equation}
 \nonumber \frac{b_1}{a_0-a_m}...\frac{b_m}{a_{m-1}-a_m}e^{-a_mt}\\
\end{equation}
\begin{equation}
\nonumber =\frac{b_m}{a_m-a_0}[\frac{b_1}{a_1-a_0}...\frac{b_{m-1}}{a_{m-1}-a_0}]e^{-a_0t}+...+\\
\end{equation}
\begin{equation}
\nonumber \frac{b_m}{a_m-a_{m-1}}[\frac{b_1}{a_0-a{m-1}}...\frac{b_{m-1}}{a_{m-2}-a_{m-1}}]e^{-a_{m-1}t}\\
\end{equation}
\begin{equation}
\nonumber +\frac{b_1}{a_0-a_m}... \frac{b_m}{a_{m-1}-a_m}e^{-a_mt}\\
\end{equation}
\begin{equation}\label{eq:App:sol6}
  = \sum_{i=0}^m \left( \prod_{k=1}^mb_k \right) \left( \prod_{j=0,j\neq i}^m\frac{1}{a_j-a_i} \right) e^{-a_it}.
\end{equation}
We note that the above expression works in the entire range of $m=0,...,N$. In the expression for $p_N(t)$, one should use $a_N=0$. The coefficients $a_i$ and $b_i$ are defined above Eq.~(\ref{eq:App:kin1}) with $m\rightarrow i$.

\subsection{Coefficient $C_m^m$}\label{sec:app:A3}

This section presents a proof of Eq.~(\ref{eq:App:sol5}) based on the  theory of functions of a complex variable.
It is recommended that the reader unfamiliar with complex analysis skips
this Section or studies basic concepts of complex integration~\cite{brown2009complex} before reading this section.

To demonstrate that the expression (\ref{eq:App:sol5}) for $C_m^m$ is valid, we show that Eq.~(\ref{eq:App:sol5}) leads to the correct initial condition $p_{m}(t=0)=\delta_{m,0}$, where $\delta_{m,0}$ is the Kronecker delta. For this purpose, it is sufficient to verify that the
right-hand side of Eq.~(\ref{eq:App:sol6}) is zero at $t=0$ for any $m>0$. Explicitly, based on Eq. (\ref{eq:App:sol6}), it is necessary to show that
\begin{eqnarray}\label{eq:App:sol7}
 0&=& \frac{1}{(a_1-a_0)(a_2-a_0)...(a_m-a_0)}+... \nonumber \\
  &+&\frac{1}{(a_0-a_m)(a_1-a_m)...(a_{m-1}-a_m)}.
\end{eqnarray}

\begin{figure}[tb]
  \centering
   \includegraphics[width=0.6\columnwidth]{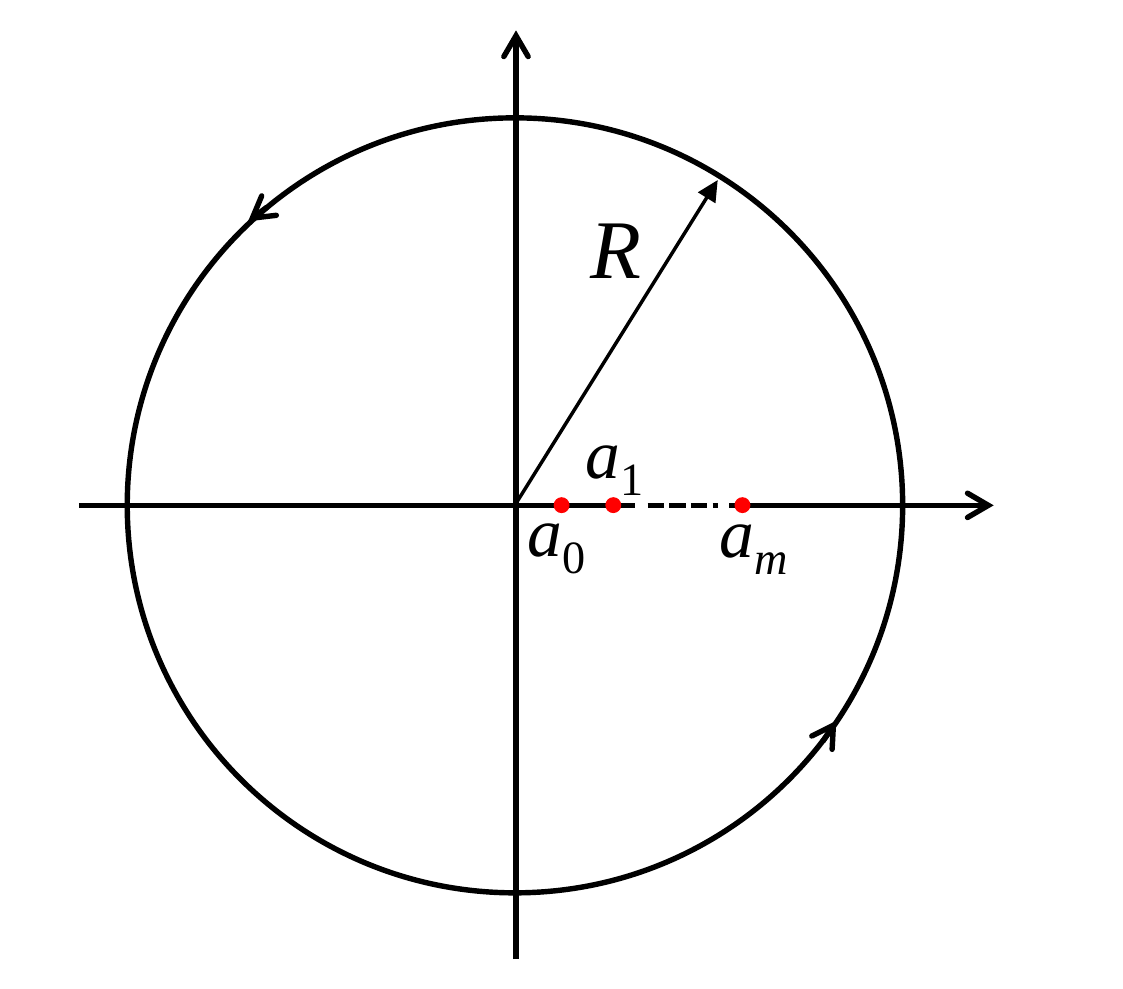}
\caption{ Path of contour integration and location of poles.} \label{fig:path}
\end{figure}

For this purpose consider a contour integral (an integral along a path in the complex plane)
\begin{equation}\label{eq:App:sol8}
  \oint\frac{1}{(a_0-z)(a_1-z)...(a_m-z)}\textnormal{d}z\equiv \oint f(z) \textnormal{d}z
\end{equation}
over a circular path $R\rightarrow \infty$, see Fig.~\ref{fig:path}. On the one hand,
it is clear that the integral is zero for $m>0$ as the modulus of integrand behaves as $1/R^{m+1}$.
 On the other hand, its value can be found using the
residue theorem~\cite{brown2009complex}. According to the residue theorem,
\begin{equation}\label{eq:App:sol8}
   \oint f(z) \textnormal{d}z=2\pi i\sum\limits_{k=0}^m \textnormal{Res}f (z),
\end{equation}
where the sum is taken over $m+1$ singularities of $f(z)$ that are $z=a_k$. Assuming that
all singularities are simple poles (as represented in Fig.~\ref{fig:path}), the residues are easily evaluated with the help of
$$
\textnormal{Res}_{z=a_k}f(z)=\left[(z-a_k)f(z)\right]\big|_{z=a_k}.
$$
The combination of these approaches (direct integration and residue theorem) leads to the relation (\ref{eq:App:sol7}).

\subsection{Average switching time}

The average switching time $ \left< T_N\right>$  into the final state $N$ can be evaluated following Eq.~(\ref{eq:T2}) approach. In the case of $N$ memristor network this time can be expressed as
\begin{equation}\label{eq:App:sol10}
  \left< T_N\right>=\int\limits_0^\infty t \frac{\textnormal{d}p_N(t)}{\textnormal{d}t} \textnormal{d}t=\int\limits_0^\infty t b_Np_{N-1}(t) \textnormal{d}t.
\end{equation}
The substitution of Eq.~(\ref{eq:App:sol6}) into Eq.~(\ref{eq:App:sol10}) results in
\begin{equation}\label{eq:App:sol11}
  \left< T_N\right>= \sum_{i=0}^{N-1}\frac{1}{a_i^2}\left( \prod_{k=1}^Nb_k \right)\left(\prod_{j=0,j\neq i}^{N-1}\frac{1}{a_j-a_i} \right).
\end{equation}

Eq.~(\ref{eq:App:sol11}) can be substantially simplified (the reader unfamiliar with complex analysis can skip this paragraph). For this purpose, we considered
a contour integral
\begin{equation}\label{eq:App:sol12}
  \oint\frac{1}{z(a_0-z)(a_1-z)...(a_N-z)}\textnormal{d}z
\end{equation}
over a circular path (as in Fig.~\ref{fig:path}) in the limit of $R\rightarrow\infty$. The evaluation of Eq. (\ref{eq:App:sol12}) integral was performed based on the residue theorem (similarly to Sec.~\ref{sec:app:A3}). On the one hand the integral is zero, on the other hand it can be calculated as a sum of all simple residues at the points $a_0, \ldots, a_{N-1}$, and the second order pole at $z=0$.
This procedure leads us to the relation
\begin{eqnarray}\label{eq:App:ResidueTrick}
  \nonumber\sum_{j=0}^{N-1}\frac{1}{a_j^2}\prod_{i=0,i\neq j}^{N-1}\frac{1}{a_i-a_j} = \\
  \nonumber =\textnormal{Res}\left(\frac{1}{(a_0-z)...(a_{N-1}-z)z^2},0\right),
\end{eqnarray}
 that was used for a simplification of the sum in Eq. (\ref{eq:App:sol11}).

Eventually, the following relation for the average switching time has been derived:
\begin{equation}\label{eq:App:sol13}
  \left< T_N\right>=\sum_{j=0}^{N-1}\frac{1}{a_j}.
\end{equation}

\section{Switching of $N$ memristors connected in-parallel} \label{app:2}

\setcounter{equation}{0}
\renewcommand\theequation{B.\arabic{equation}}

Consider the dynamics of $N$ identical probabilistic memristors connected in-parallel to a constant voltage source $V_a$ (in the past, a similar
circuit configuration was considered by Molter and Nugent~\cite{molter2016generalized}).
It is assumed that at $t=0$ all memristors are in the off-state, and the applied voltage induces their switching into the on-state.
The dynamics of each memristor is given by the following kinetic equation
\begin{equation}\label{eq:ap2:1}
  \frac{\textnormal{d}p_{0}(t)}{\textnormal{d}t} = -\gamma_{0}^1p_{0},
\end{equation}
whose solution
\begin{equation}\label{eq:ap2:2}
  p_{0}(t)=e^{-\gamma_{0}^1t}
\end{equation}
gives the probability to find the memristor in the off-state, while the probability to find it in the on-state is
\begin{equation}\label{eq:ap2:2}
  p_{1}(t)=1-e^{-\gamma_{0}^1t}.
\end{equation}

As memristors connected in-parallel are independent, the probability to find the system with all memristors in the on-state is given by the product of individual probabilities, namely,
\begin{equation}\label{eq:ap2:3}
  p_{11...11}(t)=p_1^N=\left(1-e^{-\gamma_{0}^1t}\right)^N.
\end{equation}
The corresponding switching time can be evaluated using
\begin{equation}\label{eq:ap2:4}
 \langle T_{\|,N}\rangle=\int\limits_0^1 t \textnormal{d}p_{11...11}=\int\limits_0^\infty t \frac{\textnormal{d}p_{11...11}}{\textnormal{d}t}\textnormal{d}t.
\end{equation}

For $N=2$, Eq. (\ref{eq:ap2:4}) leads to
\begin{equation}\label{eq:ap2:5}
\langle  T_{\|,2}\rangle=2\int\limits_0^\infty t \left(1-e^{-\gamma_{0}^1t}\right)e^{-\gamma_{0}^1t} \gamma_{0}^1\textnormal{d}t=\frac{3}{2\gamma_{0}^1}.
\end{equation}

For an arbitrary $N$, Eq.~(\ref{eq:ap2:4}) leads to
\begin{equation}\label{eq:ap2:6}
\langle T_{\|,N}\rangle=\frac{1}{\gamma_{0}^1}\left(1+ \frac{1}{2}+ \frac{1}{3}+...+ \frac{1}{N} \right).
\end{equation}
The above equation is the exact expression for the average switching
time of $N$ memristors connected in-parallel. The asymptotic behavior of (\ref{eq:ap2:6}) at $N\rightarrow\infty$ can be understood from the following expression, which is well-known:
\begin{equation}\label{eq:ap2:7}
  \sum\limits_{k=1}^N\frac{1}{k}=\ln N+\gamma +\mathcal{O}\left(\frac{1}{N}\right),
\end{equation}
where $\gamma\approx 0.577$ is Euler's constant. It is convention to use $\gamma$ to denote Euler's constant, but it has no relation to the $\gamma$'s used to define the switching rates.

We note that all formulae for $N$ memristors connected in-parallel can be obtained from expressions in Appendix~\ref{app:1} in the limit of equal transition rates.

\section{Some relations related to the joint switching probability distribution $\Phi(t_1,t_2)$} \label{app:3}

\setcounter{equation}{0}
\renewcommand\theequation{C.\arabic{equation}}

It is straightforward to derive the following results based on Eq.~(\ref{eq:joint_p}) for $\Phi(t_1,t_2)$. The average network switching time for the case $N=2$ can be calculated as
\begin{equation}\label{eq:T11}
\langle  T_{11}\rangle =\int\limits_0^\infty
\int\limits_0^\infty \textnormal{d}t_1
\textnormal{d}t_2
 \max(t_1,t_2)
 \Phi(t_1,t_2),
\end{equation}
where the function $\max(t_1,t_2)$ returns the maximum of two switching times $t_1$ and $t_2$.  This definition leads to the same result~Eq.(\ref{eq:T2}) that is based on the kinetic equation approach. Indeed, Eq.~(\ref{eq:T11}) can be rewritten as
\begin{equation} \label{eq:T11a1}
\int\limits_0^\infty\int\limits_0^{t_1}  t_1 \Phi(t_1,t_2)\textnormal{d}t_2
\textnormal{d}t_1 + \int\limits_0^\infty\int\limits_0^{t_2}  t_2 \Phi(t_2,t_1)\textnormal{d}t_1
\textnormal{d}t_2,
\end{equation}
where, according to the note below Eq.~(\ref{eq:joint_p}), we interchanged $t_1$ and $t_2$ in the second switching probability distribution function that corresponds to $t_1<t_2$. As both terms in the expression (\ref{eq:T11a1}) are apparently the same, one gets
\begin{eqnarray*}
\langle T_{11}\rangle = 2\gamma_{00}^1\gamma_{01}^2\int\limits_0^\infty\int\limits_0^{t_1}t_1e^{-2\gamma_{00}^1t_2}e^{-\gamma_{01}^2t_1}e^{\gamma_{01}^2t_2}\textnormal{d}t_2\textnormal{d}t_1 \\
=\frac{2\gamma_{00}^1\gamma_{01}^2}{2\gamma_{00}^1-\gamma_{01}^2}\int\limits_0^\infty t_1\left(e^{-\gamma_{01}^2t_1}-e^{-2\gamma_{00}^1t_1}\right)\textnormal{d}t_1
=\frac{1}{2\gamma_{00}^1}+\frac{1}{\gamma_{01}^2}.
\end{eqnarray*}

Besides,  the switching probability distribution $\Phi_1(t_1)$ can be calculated by integration with respect to another switching time $t_2$:
\begin{equation}\label{eq:F1}
\Phi_1(t_1) =\int\limits_0^\infty
 \Phi(t_1,t_2)\textnormal{d}t_2.
\end{equation}
Moreover, as $p_{11}$ is the probability of finding both memristors switched at the moment time $t$, it coincides with the probability of the switching of the first and the second memristor somewhere within the time interval $(0,t)$. It means that the moments of times $t_1$ and $t_2$ of the switching of the first and the second memristors should belong to the same interval, $0<t_1<t$ and  $0<t_2<t$. Therefore, by using the definition of the joint probability distribution function (see Eq.~(\ref{eq:joint_p}) and the paragraph after it), the probability $p_{11}$ can be calculated as the double integral over the square area $0<t_{1,2}<t$ of the function $\Phi(t_1, t_2)$:
\begin{equation}\label{eq:p11}
p_{11}(t) =\int\limits_0^t
\int\limits_0^t
\Phi(t_1,t_2)\textnormal{d}t_2\textnormal{d}t_1.
\end{equation}
The same idea can be used to calculate the other probabilities $p_{01}$, $p_{00}$. The result is
\begin{equation}\label{eq:p01}
p_{01}(t) =\int\limits_0^t
\int\limits_t^\infty
 \Phi(t_1,t_2) \textnormal{d}t_2\textnormal{d}t_1,
\end{equation}
and
\begin{equation}\label{eq:p00}
p_{00}(t) =\int\limits_t^\infty
\int\limits_t^\infty
\Phi(t_1,t_2)\textnormal{d}t_2\textnormal{d}t_1.
\end{equation}

\section{Adaptive probabilistic threshold model (APTM)} \label{app:4}

\setcounter{equation}{0}
\renewcommand\theequation{D.\arabic{equation}}

The threshold-type resistance switching models~\cite{pershin09b,kvatinsky2012team,Kvatinsky15a} have gained popularity and significant research
is being carried out based on such models. Here we formulate an adaptive probabilistic threshold model for probabilistic memristor modeling.

Similarly to the main text, we consider binary memristors characterized by two possible resistance states, $R_{on}$ and $R_{off}$, which can be voltage-dependent. The following voltage-dependent switching rates are postulated:
 \begin{equation}
 \gamma_{0\rightarrow 1}(V)=\left\{
 \begin{array}{cl}
k_{on}\left( \frac{V}{V_{on}}-1 \right)^{\alpha_{on}},&V> V_{on}>0 \\
0,& \textnormal{otherwise} \\
\end{array}\right.\ \label{eq:APTM1}
\end{equation}
\begin{equation}
 \gamma_{1\rightarrow 0}(V)=\left\{
 \begin{array}{cl}
k_{off}\left( \frac{V}{V_{off}}-1 \right)^{\alpha_{off}},&V< V_{off} < 0 \\
0,& \textnormal{otherwise} \\
\end{array}\right.\ \label{eq:APTM2}
\end{equation}
 where $V_{on}$ and $V_{off}$ are the threshold voltages for the transition into the off- and on-states, respectively, $k_{on}$, $k_{off}$,
 $\alpha_{on}$, and $\alpha_{off}$ are constants. As above, the probability of a memristor in the off-state at time $t$ to switch within the time interval $t$ to $t+\Delta t$ is $\Delta t \gamma_{0\rightarrow 1}(V)$. For a memristor in the on-state the probability is $\Delta t \gamma_{1\rightarrow 0}(V)$.
These switching rates, in combination with Eq. (1), take into account a wide range of possible switching behaviors.

\begin{figure}[h]
  \centering
   \includegraphics[width=0.75\columnwidth]{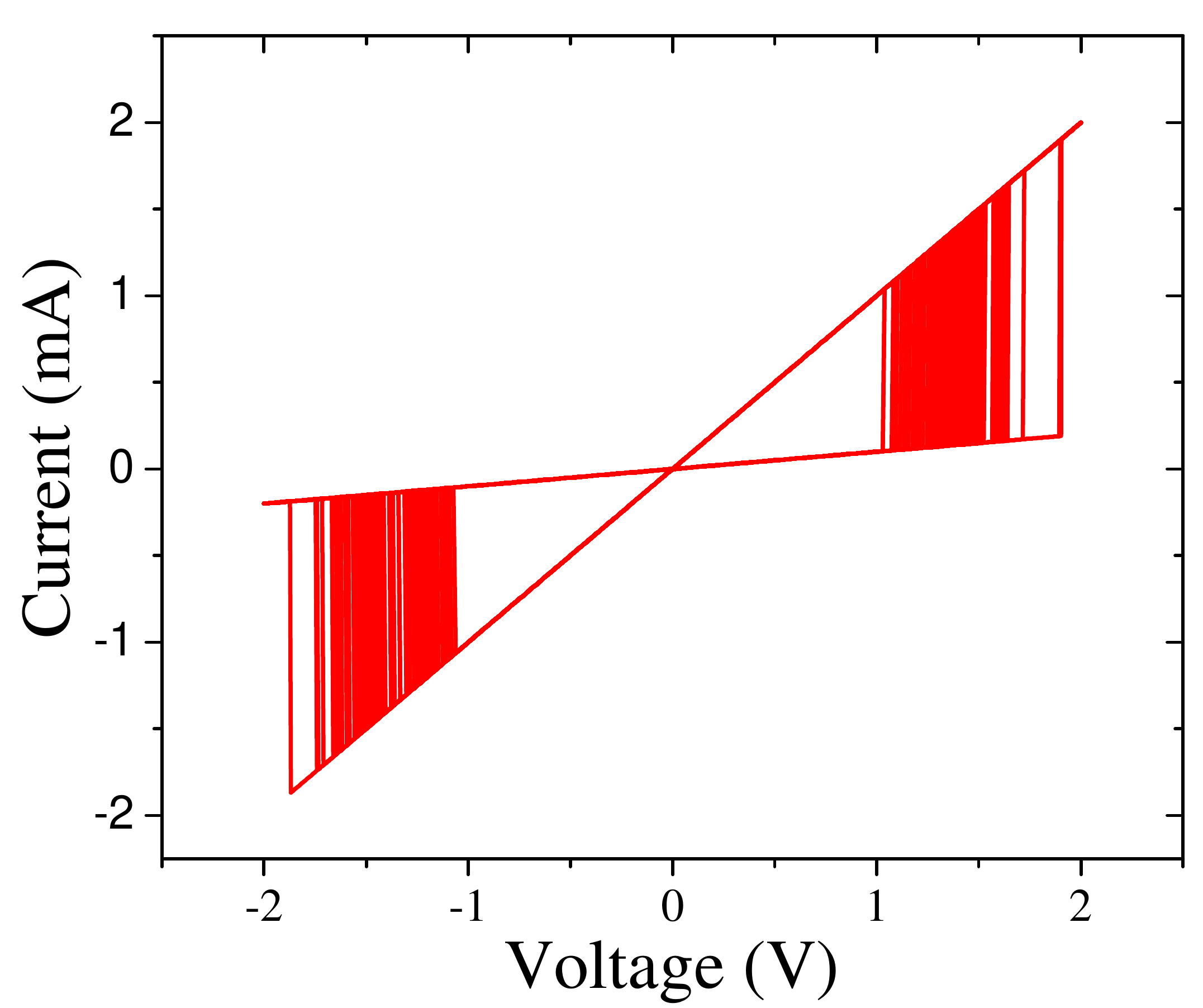}
\caption{$I-V$ curve for APTM memristor driven by a $2$~V amplitude $1$~kHz sinusoidal voltage. This plot was obtained using the following parameter values:
$k_{on}=k_{off}=10^5$~Hz, $V_{on}=1$~V, $V_{off}=-1$~V, $\alpha_{on}=\alpha_{off}=1$, $R_{on}=1$~k$\Omega$, and $R_{off}=10$~k$\Omega$.
} \label{fig:ap:2}
\end{figure}

An example of current-voltage characteristics for APTM memristor is presented in Fig.~\ref{fig:ap:2}.

\bibliographystyle{elsarticle-num}
\bibliography{memcapacitor}
\end{document}